\documentclass[twocolumn,showpacs,preprintnumbers,amsmath,amssymb]{revtex4}
\usepackage{graphicx}% Include figure files
\usepackage{dcolumn}% Align table columns on decimal point
\usepackage{bm}% bold math
\usepackage{epsfig}

\begin{document}

%\preprint{nucl-ex/oops}

\title{
Investigation of the conjectured nucleon deformation at low momentum transfer.
}

\author{N.F.~Sparveris$^1$, R.~Alarcon$^2$,  A.M.~Bernstein$^3$, W.~Bertozzi$^3$,
T.~Botto$^{1,3}$, P.~Bourgeois$^4$, J.~Calarco$^5$,
F.~Casagrande$^3$, M.O.~Distler$^6$, K.~Dow$^3$, M.~Farkondeh$^3$,
S.~Georgakopoulos$^1$, S.~Gilad$^3$, R.~Hicks$^4$, M.~Holtrop$^5$,
A.~Hotta$^4$, X.~Jiang$^4$, A.~Karabarbounis$^1$,
J.~Kirkpatrick$^5$, S.~Kowalski$^3$, R.~Milner$^3$,
R.~Miskimen$^4$, I.~Nakagawa$^3$,
C.N.~Papanicolas$^1$\footnote{corresponding author, Email address:
cnp@iasa.gr}, A.J.~Sarty$^7$, Y.~Sato$^8$, S.~\v{S}irca$^3$,
J.~Shaw$^4$, E.~Six$^2$, S.~Stave$^3$, E.~Stiliaris$^1$,
T.~Tamae$^8$,G.~Tsentalovich$^3$, C.~Tschalaer$^3$,
W.~Turchinetz$^3$, Z.-L.~Zhou$^3$ and~T.~Zwart$^3$}

\affiliation{$^1$Institute of Accelerating Systems and Applications and Department of Physics, University of Athens, Athens, Greece}
\affiliation{$^2$Department of Physics and Astronomy, Arizona State University, Temple, Arizona 85287}
\affiliation{$^3$Department of Physics, Laboratory for Nuclear Science and Bates Linear Accelerator Center,
         \\ Massachusetts Institute of Technology, Cambridge, Massachusetts 02139}
\affiliation{$^4$Department of Physics, University of Massachusetts, Amherst, Massachusetts 01003}
\affiliation{$^5$Department of Physics, University of New Hampshire, Durham, NH 03824}
\affiliation{$^6$Institut fur Kernphysik, Universitaet Mainz, Mainz, Germany}
\affiliation{$^7$Department of Astronomy and Physics, St. Mary's University, Halifax, Nova Scotia, Canada}
\affiliation{$^8$Laboratory of Nuclear Science, Tohoku University, Mikamine, Taihaku-ku, Sendai 982-0826, Japan}

\date{\today}

\begin{abstract}
We report new precise H$(e,e^\prime p)\pi^0$ measurements at the
$\Delta(1232)$ resonance at $Q^2= 0.127$ (GeV/c)$^2$ using the
MIT/Bates out-of-plane scattering (OOPS) facility.  The data reported
here are particularly sensitive to the transverse electric
amplitude ($E2$) of the $\gamma^* N\rightarrow\Delta$ transition.
Analyzed together with previous data
yield precise quadrupole to dipole amplitude ratios
$EMR = (-2.3 \pm 0.3_{stat+sys} \pm 0.6_{model})\%$ and
$CMR = (-6.1 \pm 0.2_{stat+sys}\pm 0.5_{model})\%$
 and for
$M^{3/2}_{1+} = (41.4 \pm 0.3_{stat+sys}\pm 0.4_{model})(10^{-3}/m_{\pi^+})$.
They  give credence to the conjecture of deformation in hadronic
systems favoring, at low $Q^2$, the dominance of mesonic effects.

\end{abstract}

%\pacs{}

\maketitle

The conjectured deviation of hadron
shapes from sphericity \cite{gla79} is the subject
of numerous experimental
\cite{pho1,pho2,bart,joo,frol,pos01,merve,kun00,Buuren,ware,spaprc}
and theoretical \cite{sato,mai00,kama,dmt00,multi,azna,said,dina}
investigations. The signature of the deformation of the nucleon
is most often sought
through the isolation of resonant quadrupole amplitudes in the
$\gamma^* N\rightarrow \Delta$ transition \cite{rev2,rev3}. The
origin of the deformation is attributed to different effects depending
on the theoretical approach adopted. In the constituent-quark
picture of the nucleon, a quadrupole resonant amplitude would
result from a d-state admixture in the 3-quark wave function, a
consequence of the color-hyperfine interaction among quarks.
In dynamic models of the $\pi N$ system, the presence of the pion
cloud gives rise to quadrupole amplitudes which dominate in the
low $Q^2$ region \cite{sato,dmt00}. At $Q^2= 0.127$
(GeV/c)$^2$ where the reported measurements have been performed,
the pionic contribution is predicted to be maximal.

Spin-parity selection rules in the $N(J^{\pi}=1/2^+)\rightarrow
\Delta(J^{\pi}=3/2^+)$ transition, allow only magnetic dipole
$(M1)$ and electric quadrupole $(E2)$ or Coulomb quadrupole
$(C2)$ photon absorption multipoles (or the corresponding pion
production multipoles $M^{3/2}_{1+}$, $E^{3/2}_{1+}$ and
$S^{3/2}_{1+}$, respectively) to contribute. The ratios
$CMR = Re(S^{3/2}_{1+}/M^{3/2}_{1+})$ and $EMR =
Re(E^{3/2}_{1+}/M^{3/2}_{1+})$ are routinely used to present the relative
magnitude of the amplitudes of interest.

The cross section of the H$(e,e^\prime p)\pi^0$ reaction is
sensitive to four independent partial cross sections
($\sigma_{T},\sigma_{L},\sigma_{LT}$ and $\sigma_{TT}$)
proportional to the corresponding response functions \cite{multi}
:

\begin{eqnarray}
 \frac{d^5\sigma}{d\omega d\Omega_e d\Omega^{cm}_{pq}} & = & \Gamma (\sigma_{T} + \epsilon{\cdot}\sigma_L
  - v_{LT}{\cdot}\sigma_{LT}{\cdot}\cos{\phi_{pq}} \nonumber \\
  &  &   +\epsilon{\cdot}\sigma_{TT}{\cdot}\cos{2\phi_{pq}} ) \nonumber
\label{equ:cros}
\end{eqnarray}
where the kinematic factor $v_{LT}=\sqrt{2\epsilon(1+\epsilon)}$
and $\epsilon$ is the transverse polarization of the virtual
photon, $\Gamma$ the virtual photon flux and $\phi_{pq}$ is the
proton azimuthal angle with respect to the momentum transfer
direction.

The $E2$ and $C2$ amplitudes manifest themselves most
prominently through interference with the dominant dipole $(M1)$
amplitude. The interference of the $C2$ amplitude with the $M1$
leads to Longitudinal~-~Transverse (LT) response while the
interference of the $E2$ amplitude with the $M1$ leads to
Transverse~-~Transverse (TT) responses. The
$\sigma_{o}=\sigma_{T}$+$\epsilon\cdot\sigma_{L}$ partial cross
section is dominated by the $M_{1+}$ multipole.

$E2$ and $EMR$ are more difficult to isolate in electro-pion production
than $C2$ and $CMR$ because the transverse responses are
dominated by the $| M_{1+} | ^2 $ term which is of course
absent in the longitudinal sector. As a result the precision with
which both $EMR$ and $CMR$ have been determined in previous
measurements is limited due to the poor determination of $EMR$ and
the correlation in the $EMR$ and $CMR$ extraction \cite{vellthes}. In
order to address this difficulty and to access $E2$ ($EMR$) with the
highest precision we have defined \cite{papa99} the partial cross
section $\sigma_{E2}$ which was measured for the first time in
this experiment. $\sigma_{E2}(\theta^*_{pq})$ is defined as:
$$\sigma_{E2}(\theta^*_{pq}) \equiv \sigma_o(\theta^*_{pq})+\sigma_{TT}(\theta^*_{pq})-\sigma_o(\theta^*_{pq}=0^0)$$
$\sigma_{E2}$  exhibits far greater sensitivity to the $EMR$ compared to the
$\sigma_{TT}$; this becomes obvious in a multipole expansion of
$\sigma_{E2}$ up to S and P waves:

$$\sigma_{E2} = 2 Re [E^*_{o+}(3E_{1+} + M_{1+} - M_{1-})](1-cos \theta^*_{pq}) $$
             $  \qquad \qquad \qquad - 12 Re [E^*_{1+}(M_{1+}-M_{1-})] sin^2 \theta^*_{pq} $

$$\sigma_{TT} = 3 sin^2 \theta^*_{pq} [ \frac{3}{2} | E_{1+} | ^2 - \frac{1}{2} | M_{1+} | ^2 \qquad \qquad \qquad $$
             $  \qquad \qquad \qquad - Re \{ E^*_{1+}[ M_{1+} - M_{1-} ] + M^*_{1+}M_{1-} \} ] $

\qquad

The $|M_{1+}|^2$ term which dominates the $\sigma_{TT}$ and
suppresses the sensitivity to the $Re(E^{*}_{1+}M_{1+})$ term at
$\theta^{*}_{pq}=90^0$ is cancelled out in the $\sigma_{E2}$
while at the same time the $Re(E^{*}_{1+}M_{1+})$ term is
magnified by a factor 12. This is the reason that lead us to
perform our measurements at $\theta^{*}_{pq}=90^0$. The very
definition of $\sigma_{E2}$ clearly shows that its experimental
determination presents formitable challenges including the issue
of containment of the systematic error.

The measurements reported here were performed using the technique
of the out-of-plane detection \cite{cnp-oops} with the Out-Of-Plane
Spectrometer (OOPS) \cite{oopsdolf,oopsmand,oopsnim} system. Three
identical OOPS modules \cite{oopsdolf,oopsmand,oopsnim} were
placed symmetrically at azimuthal angles $\phi_{pq}= 60^0$, 90$^0$
and $180^0$ with respect to the momentum transfer direction for
the measurement at central kinematics of $\theta_{pq}^{*}$=$90^0$
and thus we were able to isolate $\sigma_{TT}$, $\sigma_{LT}$ and
$\sigma_{o}=\sigma_{T}$+$ \epsilon \cdot \sigma_{L}$. An OOPS
spectrometer module was also positioned along the momentum
transfer direction. This measurement provided directly the
parallel cross section determination
$\sigma_{o}(\theta_{pq}^{*}=0^0)$. Measurements were taken at $W$
= 1232 MeV, $Q^2$ =  0.127 GeV$^2$/$c^2$ and central proton
angles of $\theta_{pq}^{*}$=$0^0$ and $90^0$ while the extensive
phase space coverage of the spectrometers allowed the extraction
of the responses at $\theta_{pq}^{*}$=$85^0$, $90^0$ and $95^0$.
It completes a series of earlier $N
\rightarrow \Delta$ Bates measurements \cite{merve,kun00,spaprc}
and is the first one to measure the $\sigma_{TT}$ and
$\sigma_{E2}$, the partial cross sections sensitive to the
Electric Quadrupole amplitude $E2$.

The experiment was performed in the South Hall of M.I.T.-Bates
Laboratory. A high duty factor 950 MeV unpolarized electron beam
was employed on a cryogenic liquid-hydrogen target. The beam
average current was 7 $\mu A$. Electrons were detected with the
OHIPS spectrometer \cite{xia98} and protons were detected with
the OOPS spectrometers \cite{oopsdolf,oopsmand,oopsnim},
symmetrically positioned with respect to the momentum transfer
direction. The OHIPS spectrometer employed two Vertical Drift
Chambers for the track reconstruction. Two layers of 18 Pb-Glass
detectors and a Cherenkov detector were responsible for
identification of electrons from the $\pi^{-}$ background. The
timing information for OHIPS derived from 3 scintillator
detectors. The OOPS spectrometers used three Horizontal Drift
Chambers for the track reconstruction followed by three
scintillator detectors for timing and for the separation of the
protons from the strong $\pi^{+}$ background coming from the
$\gamma^* p \rightarrow \pi^{+} n$ process. The uncertainty in
the determination of the central momentum was 0.1\% for the
proton arm and 0.15\% for the electron arm. The spectrometers
were aligned with a precision better than 1 mm and 1 mrad, while
the uncertainty and the spread of the beam energy were $0.3\%$
and $0.03\%$ respectively. An OOPS spectrometer was used
throughout the experiment as a luminosity monitor detecting
elastically scattered protons. A detailed description of all
experimental uncertainties and their resulting effects in the
measured responses is presented in \cite{sparve}.

Elastic scattering data for calibration purposes were taken using
liquid-hydrogen and carbon targets and a 600 MeV beam.
Measurements with and without sieve slits for all spectrometers
allowed the determination of the optical matrix elements for all
spectrometers and their absolute efficiency.
The consistency of the new with the previous measurements
\cite{merve} is confirmed  through their excellent agreement in
the parallel cross section $\sigma_{o}(\theta_{pq}^{*}=0^0)$.

\begin{figure*}
\centerline{\psfig{figure=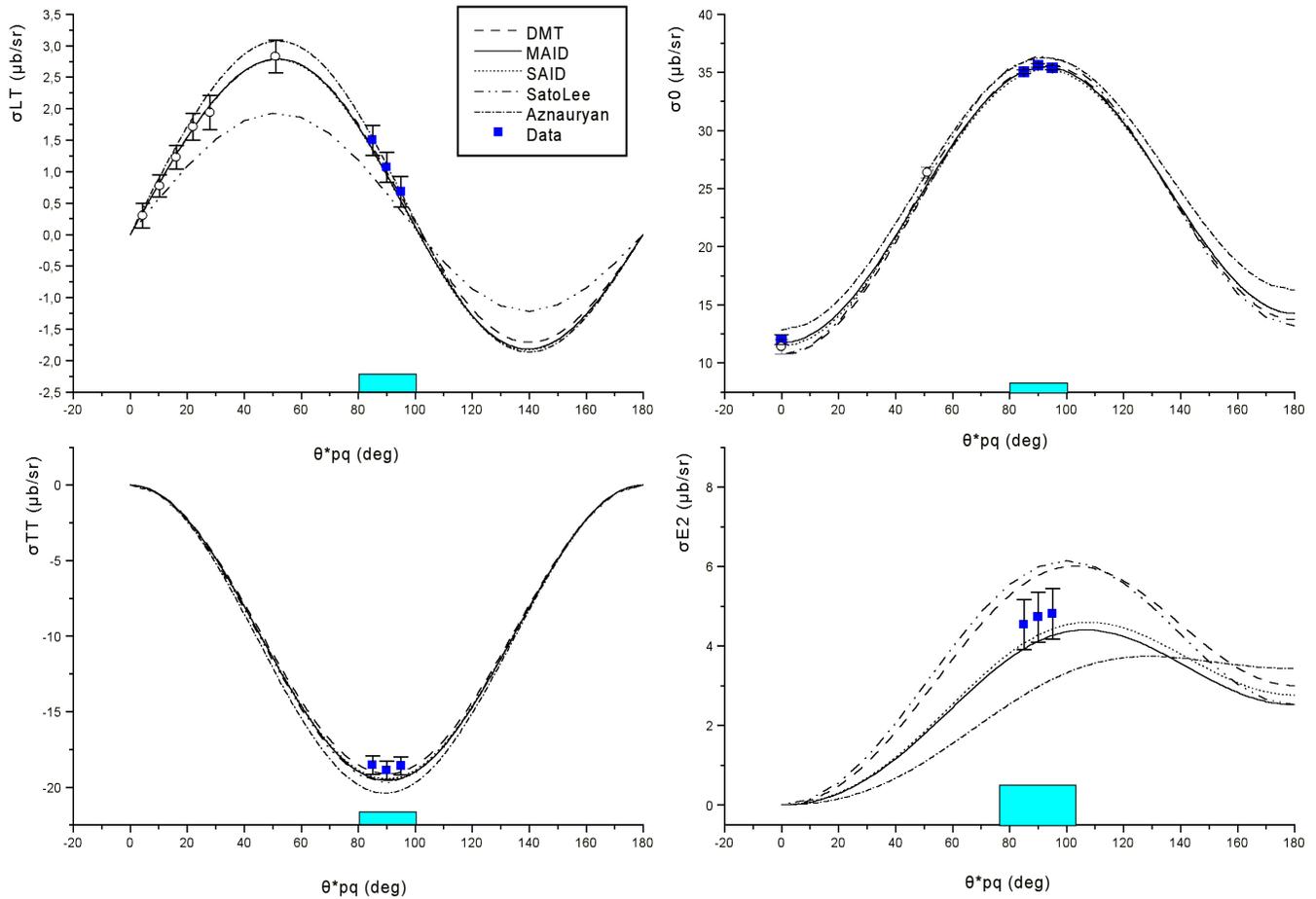,width=15.0cm,angle=270.0}}
\smallskip
\caption{The measured $\sigma_{LT}$, $\sigma_{TT}$, $\sigma_{o} =
\sigma_{T} + \epsilon \cdot \sigma_{L}$ and $\sigma_{E2}$ partial
cross sections as a function of $\theta^*_{pq}$. Solid square
points depict this experiment's results while the open circle
points correspond the results from the previous
Bates experiments \cite{merve,kun00}.
%\cite{merve,kun00}.
The shaded bands depict the corresponding
systematic uncertainty.} \label{fig:rlt}
\end{figure*}

In Figure~\ref{fig:rlt} we present the experimental results for
$\sigma_{TT}$, $\sigma_{LT}$,
$\sigma_{T}$+$\epsilon{\cdot}\sigma_{L}$ and $\sigma_{E2}$ along
with those of earlier Bates experiments \cite{merve,kun00}. They
are compared with the SAID multipole analysis \cite{said}, the
phenomenological model MAID 2000 \cite{mai00,kama}, the Aznauryan
dispersion analysis \cite{azna}, the Dynamical Models of Sato-Lee
(SL) \cite{sato} and of DMT (Dubna - Mainz - Taipei) \cite{dmt00}.
Results from these models have been widely used in comparisons
with recent experimental results
\cite{pho1,pho2,bart,joo,frol,pos01}; a description of their physical content is presented in the original papers.

%-----------------------------------------------------------------------------
%-----------------------------------------------------------------------------

The SAID multipole analysis \cite{said} is capable of successfully describing the new data,
as can be seen in Figure~\ref{fig:rlt}, but not the corresponding recoil
polarization data \cite{pos01,ware}. The data basis at $Q^2$ =  0.127 GeV$^2$/c$^2$ is found to
be not rich enough to provide a stable solution. It
is hoped that the addition of the $H(e,e^\prime \pi^+)$ data
(same $Q^2$) which are now being analyzed will
provide sufficiently rich basis for an independent solution.

The MAID model \cite{mai00,kama} which offers a flexible
phenomenology also provides a successful description of the new
data especially when its parameters are re-adjusted. In addition
it offers consistently good description of all available
measurements \cite{merve,kun00,ware} at this $Q^2$; it is the only
model that succeeds in this very demanding task.

The fixed $t$ dispersion analysis applied by I.~Aznauryan
\cite{azna} provides an alternative phenomenological approach to
describing the data and extracting the multipole information of
interest. It is also able to fit the new data remarkably well; the
same holds for the case of earlier measurements at the same $Q^2$
but it still disagrees with our $\sigma_{LT}$  measurement at
$W$=1170 MeV \cite{kun00} .

The SL \cite{sato} and DMT \cite{dmt00} dynamical models provide a
nucleon description which incorporates physics of the pionic
cloud. Both calculate the resonant channels from dynamical
equations. DMT uses the background amplitudes of MAID with some
small modifications.  SL calculate all amplitudes consistently
within the same framework with only three free
parameters. Both models predict that large fraction of the $E2$
and $C2$ multipole strength arise due to the pionic cloud with the
effect reaching a maximum value in the region of $Q^2$ of our
measurements. The SL model disagrees with our
$\sigma_{LT}$ measurements but also with our earlier
$\sigma_{LT'}$ and polarization results \cite{kun00}. DMT offers a
good agreement with our data at resonance but it fails to describe
$\sigma_{o}$ and $\sigma_{LT}$ below resonance as well as the W
dependence of the parallel cross section \cite{kun00,spaprc}. The
results of both SL and DMT taken together indicate that the
dynamical models offer a promising phenomenology for exploring the
role of the pionic cloud to the issue of deformation, but they
have not yet achieved a satisfactory description of the data in
the region where the pionic cloud effect is expected to be
maximal.

In Table~\ref{tab:resu} the resonant $M_{1+}(3/2)$ and $CMR$ and $EMR$
derived or used by the aforementioned models are listed along with
the results from a Truncated Multipole Expansion (TME) fit  to our
data \cite{sparve}. In the TME fit, as in \cite{merve}, it is assumed
that only the resonant amplitudes
($M^{3/2}_{1+}$, $E^{3/2}_{1+}$ and $S^{3/2}_{1+}$)
contribute. As documented in \cite{merve,sparve_new} the results of
the TME are compatible with those of MAID if the truncation (model)
error due to the omission of
higher waves is taken into account. Table~\ref{tab:resu} shows
that an overall consistency in terms of the expected sign and
magnitude has emerged; however, a quantitative agreement has not yet been achieved.
The issue of model error not
withstanding, such a comparison is not warranted since only one
model, MAID, provides an overall agreement with the entire data
base of our results at this momentum transfer. For this reason, and in accordance
with earlier publications \cite{merve,pos01} we adopt the values derived from the
MAID fit \cite{kama03} to our data:
$M^{3/2}_{1+} = (41.4 \pm 0.3_{stat+sys}\pm 0.4_{model})(10^{-3}/m_{\pi^+})$,
$EMR = (-2.3 \pm 0.3_{stat+sys} \pm 0.6_{model})\%$ and
$CMR = (-6.1 \pm 0.2_{stat+sys}\pm 0.5_{model})\%$.

\begin{table}
\begin{ruledtabular}
\begin{tabular}{||l||lll||}
\hline
            &  $CMR (\%)$        & $EMR (\%)$      & $M^{3/2}_{1+} (10^{-3}/m_{\pi^+})$ \\
\hline
TME         &  $-6.9 \pm 0.4$    & $-3.1 \pm 0.5$  &  $41.6 \pm 0.3 $            \\
            &                    &                 &                             \\
SAID        &  $-4.8        $    & $-1.4        $  &  $39.7         $            \\
MAID        &  $-6.1 \pm 0.2$    & $-2.3 \pm 0.3$  &  $41.4 \pm 0.3 $            \\
Aznauryan   &  $-7.9 \pm 0.9$    & $-0.9 \pm 0.5$  &  $40.8 \pm 0.5 $            \\
Sato Lee    &  $-4.3        $    & $-3.2        $  &  $41.7         $            \\
DMT         &  $-6.1 \pm 0.3$    & $-1.9 \pm 0.3$  &  $41.5 \pm 0.4 $            \\
\hline
\end{tabular}
\end{ruledtabular}
\caption{Values of $CMR$ and $EMR$ and $M_{1+}$ for the SAID, MAID, Aznauryan, SL and DMT Models at $Q^2= 0.127$ (GeV/c)$^2$. The values
quoted with uncertainties result from an adjustment of the model
parameters to fit our data. The result from the Truncated
Multipole Expansion (TME) fit to the data is also presented.}
\label{tab:resu}
\end{table}

The quoted model error is a conservative estimate of the uncertainty arising from the employment of multipoles
in MAID not constrained by our measurements\cite{sparve,sparve_new,itaru}.
The new results are consistent with our earlier results \cite{merve,kun00} but significantly more accurate.
They are the most accurately known $CMR$ and $EMR$ at any finite $Q^2$ value.

In conclusion, the new data, and in particular those concerning
$\sigma_{TT}$ and the newly introduced $\sigma_{E2}$ partial cross
sections, taken together with our previous measurements
provide a precise determination of both $EMR$ and $CMR$ at
$Q^2$= 0.127 (GeV/c)$^2$. Both ratios are substantially bigger
(by an order of magnitude) than the values predicted by quark models
on account of color hyperfine interaction. They are consistent in magnitude
with the values predicted by models taking into account the mesonic
degrees of freedom. This we interpret as a validation of the crucial role
the pion cloud plays in nucleon structure, a consequence of the spontaneous
breaking of chiral symmetry \cite{rev3}. However, at this $Q^2$ where pionic effects are expected to be manifesting themselves maximally, dynamical
models \cite{sato,dmt00} fail to describe the experimental
quantities in detail. Finally, we observe that the non zero values of these resonant quadrupole amplitudes determined in this experiment confirm that the nucleon or its first excited state, the delta, or more likely both are deformed.

We are indebted and would like to thank Drs. I.~Aznauryan,
S.S.~Kamalov, T.-S.H.~Lee, T.~Sato, I.~Strakovsky and L.~Tiator
for providing us with valuable suggestions on the overall program
and these results in particular. This work is supported in part by
the US Department of Energy, the National Science Foundation, by a
Grant in Aid for Scientific research (KAKENHI, No 14540239), the
Japan Society for the promotion of Science and the EC-IHP Network
ESOP, Contract HPTN-CT-2000-00130.

\end{document}